\documentstyle[aps,prb,preprint,tighten,epsfig]{revtex}
\begin{document}
\draft 

\title{Structural, Electronic, and Magnetic Properties of MnO}
\author{J.E. Pask, D.J. Singh, I.I. Mazin, C.S. Hellberg, and J. Kortus}
\address{Center for Computational Materials Science, Naval Research
 Laboratory, Washington, DC 20375}
\date{\today}
\maketitle

\begin{abstract} 
We calculate the structural, electronic, and magnetic properties of MnO from
first principles, using the full-potential linearized augmented planewave
method, with both local-density and generalized-gradient approximations to
exchange and correlation. We find the ground state to be of rhombohedrally
distorted B1 structure with compression along the [111] direction,
antiferromagnetic with type-II ordering, and insulating, consistent with
experiment. We show that the distortion can be understood in terms of a
Heisenberg model with distance dependent nearest-neighbor and
next-nearest-neighbor interactions determined from first principles. Finally,
we show that magnetic ordering can induce significant charge anisotropy, and
give predictions for electric field gradients in the observed rhombohedrally
distorted structure.
\end{abstract}

\pacs{PACS: 75.30.Et, 75.10.Lp, 75.50.Ee, 76.60.Gv}

\section{Introduction} 
\label{intro}  

Since the recognition of their unusual insulating and magnetic properties over
a half century ago, the first-row transition-metal monoxides have been the
subject of much experimental and theoretical interest. Because they are highly
correlated Mott or charge-transfer insulators,\cite{ref1,ref2} first-principles
theoretical approaches based on density-functional (DFT) in the local-density
approximation (LDA)\cite{ref3,ref4} have proven inadequate for understanding
certain properties such as electrical conductivity and magnetic moments, and
new or extended approaches such as self-interaction corrected
(SIC),\cite{ref5,ref6} orbital polarization corrected,\cite{ref7} on-site
Coulomb corrected (LDA+U),\cite{ref8,ref9} optimized effective potential
(OEP),\cite{ref9a} and model GW\cite{ref10} methods have been developed in
order to address these deficiencies. Nevertheless, DFT-LDA based approaches
have proven successful in understanding a number of ground-state properties,
most notably in the cases of MnO and NiO which are correctly predicted to be
insulating and antiferromagnetic with type-II ordering (though the gaps are
underestimated).\cite{ref11,ref12,ref13,ref14} Moreover, the use of the
generalized gradient approximation (GGA)\cite{ref15} has been shown to yield
further improvements.\cite{ref16} Here, we focus on the larger-moment compound,
MnO, from the standpoint of DFT in both LDA and GGA approximations.

Above the N\'eel temperature, $T_N = 118$ K, MnO is a paramagnetic insulator
with the rocksalt (B1) structure. Below $T_N$, it is a type-II
antiferromagnetic (AFII, Fig.~\ref{fig1}) insulator with rhombohedrally
distorted B1 structure. The rhombohedral distortion takes the form of a
compression along the [111] direction, taking cubic angles from $90^\circ$ to
$90.62^\circ$.\cite{ref17} Based upon augmented spherical wave (ASW) LDA
calculations, Oguchi {\em et al.}\cite{ref12,ref13} showed that the electronic
structure is very sensitive to the magnetic ordering, and that the LDA can
predict MnO to be an insulator, but only when the magnetic ordering is AFII, as
it is experimentally. Based again upon ASW-LDA calculations, Terakura {\em et
al.}\cite{ref14} subsequently explained the stability of the AFII ordering
relative to nonmagnetic (NM), ferromagnetic (FM), and type-I antiferromagnetic
(AFI) orderings in terms of the strong $dd\sigma$ coupling through the oxygens
of opposite-spin cation sublattices in the AFII ordering vs.\ same-spin
sublattices in the AFI. Using a linear combination of atomic orbitals (LCAO)
LDA approach, Belkhir and Hugel\cite{ref17a} examined the splitting of {\em d}
subbands, concluding that {\em pd} hybridization in the AFII structure enhances
the splitting relative to that predicted by classical ligand field theory.
Using a linearized augmented planewave (LAPW) approach, Dufek {\em et
al.}\cite{ref16} showed that the GGA can yield further improvements to LDA
results, finding a minor improvement of the gap from 1.0 eV to 1.4 eV and of
the Mn spin magnetic moments from 3.72$\mu_B$ to 4.15$\mu_B$, relative to
experimental values of 3.6--3.8 eV and 4.58--4.79$\mu_B$, respectively. Using
an LCAO Hartree-Fock (HF) approach, Towler {\em et al.}\cite{ref18} calculated
structural, magnetic, elastic, and vibrational properties. They determined the
correct AFII ground state and a rhombohedral distortion angle of $90.47^\circ$,
very close to the experimental value. Mackrodt {\em et al.}\cite{ref19}
subsequently applied the same methodology to the calculation of phase
transitions under pressure. Using a linear muffin tin orbital atomic sphere
approximation (LMTO-ASA) approach, Cohen {\em et al.}\cite{ref20} calculated
low- and high-pressure properties of MnO in the context of an investigation of
magnetic moment collapse, and found that the GGA yielded an improved lattice
constant, bulk modulus, and magnetic moment relative to LDA results. More
recently, using a planewave pseudopotential (PWPP) GGA approach, supplemented
by LDA+U calculations, Fang {\em et al.}\cite{ref21} investigated low- and
high-pressure properties. They determined the correct AFII ground state, a
rhombohedral distortion angle of $\approx 92^\circ$, somewhat larger than the
experimental value, magnetic moments, and a phase transition to a metallic NiAs
structure at high pressure. Massidda {\em et al.}\cite{ref22} have investigated
magnetic-order-induced anisotropies in linear-response properties using a
combination of {\em ab initio} and model calculations. In the course of their
study, they determined a ground-state distortion angle of $90.66^\circ$, very
close to the experimental value, and showed that the zone-center optic phonon
frequencies and Born effective charge tensor exhibit significant
magnetic-order-induced anisotropies, even when assuming the perfect rocksalt
structure, contrary to the assumption of cubic symmetry commonly made in
interpreting experimental data.

The mechanism underlying the rhombohedral distortion below $T_N$ was examined
by Kanamori\cite{ref23} in the late 1950's. Based upon paramagnetic 
susceptibility data, it was concluded that there are significant exchange
interactions between nearest-neighbor (NN) cations and, based upon Curie-Weiss
and N\'eel temperature data, that these interactions are predominantly direct
in nature. It was further concluded, consistent with earlier suggestions of
Greenwald and Smart,\cite{ref24} that the rhombohedral distortion below $T_N$
is due to these NN interactions. Assuming distance dependent NN and
next-nearest-neighbor (NNN) interactions, Rodbell and Owen\cite{ref25}
subsequently used a molecular field approach to derive an expression for the
distortion in terms of these interactions which supported the above
conclusions, having explicit dependence only upon the NN interactions. Lines
and Jones\cite{ref26} subsequently deduced the same expression from a
Heisenberg model containing only NN and NNN interactions, and used a Green's
function approach\cite{ref27} in the random phase approximation to evaluate the
exchange constants based upon susceptibility data. More recently, Oguchi {\em
et al.}\cite{ref12,ref13} calculated exchange constants based on {\em ab
initio} ASW-LDA calculations, employing the Korringa-Kohn-Rostoker coherent
potential approximation (KKR-CPA). In this work they showed that interactions
beyond NNN were in fact negligible, justifying the assumption made almost
universally in previous work. The constants obtained, however, were about three
times larger than those determined from experimental data. Among the reasons
suggested for the discrepancy were the assumption that total-energy differences
were well represented by eigenvalue-sum differences, and spherical
approximations in the ASW and KKR-CPA calculations. Subsequently, Solovyev and
Terakura\cite{ref27a} calculated exchange constants based on LMTO-ASA-LDA
calculations in the context of a discussion of single-particle theoretical
approaches to MnO. In this work, they employed an expression\cite{ref27b} for
the exchange constants based on infinitesimal displacements from the ground
state configuration rather than finite rotations in a random medium as in
Refs.~\onlinecite{ref12} and \onlinecite{ref13}, and  employed spherical
approximations as before. The constants obtained were, however, again
significantly larger than those determined from experimental data. The main
reason suggested for the discrepancy was the underestimation of intra-atomic
exchange and charge-transfer energies in the LDA.

Here, we calculate the structural, electronic, and magnetic properties of MnO
by the full-potential linearized augmented planewave method, using both LDA and
GGA approximations to exchange and correlation, thus improving upon spherical,
pseudopotential, and/or strictly LDA approximations in earlier work. We show
that the observed rhombohedral distortion can be understood in terms of a
Heisenberg model with distance dependent NN and NNN interactions determined
from {\em ab initio} calculations, without the CPA or spherical approximations.
And lastly, we consider the charge anisotropy induced by magnetic ordering
through {\em ab initio} calculations of electric field gradients.

\section{Crystal, electronic, and magnetic structure}
\label{cryselmagstruc}

General potential LAPW\cite{ref28} calculations were carried out for NM, FM,
AFI, and AFII  magnetic orderings. Well converged basis sets were employed with
a planewave cutoff of 18.9 Ry, corresponding to an average of $\approx 109$
basis functions per atom (the exact number being {\bf
k}-dependent).\cite{ref30} Local orbital extensions were used for Mn $l=1,2,3$
and O $l=1,2$ channels to include higher lying semi-core states and relax
linearization errors. Core states were treated fully relativistically in a
self-consistent atomic-like approximation, while valence states were treated
scalar relativistically. Increasing the planewave cutoff to 24.7 Ry,
corresponding to an average of $\approx 159$ basis functions per atom, yielded
a change in total energy of less than 2.5 mRy/atom in LDA tests. Highly
converged special {\bf k} point sets\cite{ref31} were employed consisting of
512 points in the rhombohedral (NM, FM, and AFII) Brillouin zones and 486
points in the tetragonal (AFI) zone. Doubling the number of {\bf k} points in
each direction yielded a change in total energy of less than 0.15 mRy/atom in
LDA tests. LDA and GGA calculations employed the exchange-correlation
functionals of Hedin and Lundqvist,\cite{ref32} and Perdew and
Wang,\cite{ref33} respectively. As shown in Fig.~\ref{fig2}, the LDA
calculations predict the ground state to be antiferromagnetic with AFII
ordering, consistent with experiment\cite{ref17} and previous HF
calculations\cite{ref18} and density-functional calculations\cite{ref14,ref21}
employing spherical or pseudopotential approximations. The GGA results are
qualitatively similar, predicting the correct AFII ground state, but with a
lattice constant of 8.38 a.u., much improved from the LDA value of 8.16 a.u.
relative to the experimental value of 8.38 a.u.\cite{ref34}

Figures \ref{fig3} and \ref{fig4}(c) show the Brillouin zone and LDA band
structure and density of states corresponding to the ground-state AFII
ordering. The GGA results are qualitatively similar, but with a larger exchange
splitting and gap of $\approx 0.29$ Ry and $\approx 0.105$ Ry, respectively,
compared to LDA values of $\approx 0.27$ Ry and $\approx 0.075$ Ry, due to
lower occupied and higher unoccupied $d$ states relative to O $p$ states. In
both cases, the large exchange splitting and narrow $e_g$ band are sufficient
to produce an insulator, consistent with experiment\cite{ref17} and previous
calculations\cite{ref9,ref14,ref16,ref18,ref21,ref22}. Significantly, however,
both LDA and GGA calculations predict other configurations to be metallic
(Figs.~\ref{fig4}(a),(b)), with the implication that the spin-disordered
paramagnetic state would also be metallic, contrary to experiment. These
results are consistent with the characterization of MnO as a correlated
Mott-Hubbard insulator.

\section{Rhombohedral distortion}
\label{rhomdist}

Figure \ref{fig5} shows LDA results for the total energy vs. rhombohedral
strain in the ground-state AFII ordering at the experimental volume. The GGA
results are essentially the same. The rhombohedral strain is determined by a
parameter $g$ through the volume conserving strain tensor
\[
G = \left(\begin{array}{ccc}
   1+g   & g   & g   \\
   g     & 1+g & g   \\
   g     & g   & 1+g
   \end{array}\right)(1+3g)^{-1/3}.
\]
As shown in the figure, the calculations predict the ground state to be
slightly compressed ($g<0$) along the [111] direction. A least squares fit of
the results yielded minimizing strains of $g=-0.01456$ and $g=-0.01465$,
corresponding to deviations of $1.68^\circ$ and $1.69^\circ$ from the
$90^\circ$ cubic angle, for LDA and GGA calculations, respectively. A
comparison with experiment and previous calculations is shown in
Table~\ref{tab1}. We find that both LDA and GGA results overestimate the
experimental angle somewhat, consistent with recent PWPP-GGA
calculations\cite{ref21} but inconsistent with recent LAPW-LDA
calculations.\cite{ref22} On the other hand, Hartree-Fock
calculations,\cite{ref18} which produce larger gaps, underestimate the
distortion somewhat. Within simple tight-binding pictures, interatomic exchange
interactions are inversely related to the gap ($\sim t^2/I$ or $\sim t^2/U$,
where $t$ is a hopping parameter and $I$ and $U$ are on-site interactions
related to the size of the gap). As discussed below, the distortion is driven
by the variation of NN exchange with distance, so the tendency to overestimate
the distortion may reflect the incomplete treatment of Hubbard correlations and
resulting small gaps associated with LDA and GGA approximations.

The internal energy of the crystal can be partitioned into elastic and exchange
parts. The elastic part is symmetric with respect to small rhombohedral
distortions of the ideal cubic structure. Thus, it is sufficient to consider
the exchange part to understand the mechanism of the observed compression along
[111]. Within the model, the size of the distortion is just due to the
competition of the exchange part which favors distortion and the non-spin-order
dependent elastic energy, given in the harmonic approximation by the shear
elastic constant, $c_{44}$ for the non-spin-ordered case. To set the energy
scale we calculated $c_{44} = 66$ GPa using the LDA at the experimental cell
volume with a ferromagnetic ordering. Experimentally, the two terms cannot be
readily separated, but above the magnetic ordering temperature a value of
$c_{44}$ near 75 GPa is measured with very strong temperature dependence in the
region of $T_N$.\cite{elastic} We now show that the exchange part can be
understood in terms of a Heisenberg Hamiltonian with distance dependent NN and
NNN interactions determined from {\em ab initio} calculations. Let
\begin{equation}
\label{eq1}
H = \sum_{NN} J_1 {\bf S}_i \cdot {\bf S}_j + 
    \sum_{NNN} J_2 {\bf S}_i \cdot {\bf S}_j.
\end{equation}
Considering a single formula unit, the above Hamiltonian reduces to
\begin{equation}
H^{\rm FM} = (6J_1+3J_2)S^2,
\end{equation}
\begin{equation}
H^{\rm AFI} = (-2J_1+3J_2)S^2,
\end{equation}
and
\begin{equation}
H^{\rm AFII} = -3J_2 S^2,
\end{equation}
where $S$ is the magnitude of the cation spin, for FM, AFI, and AFII orderings,
respectively. We deduce the exchange constants from the above expressions and
{\em ab initio} total energies for each ordering at the experimental volume,
with $S=5/2$. Our results are shown in Table~\ref{tab2}, along with previous 
{\em ab initio}\cite{ref12,ref13,ref27a} and
semi-empirical\cite{ref26,ref35,ref36} results for comparison. While there are
still significant differences, our {\em ab initio} results agree in sign and
order of magnitude with the semi-empirical ones, and are generally closer to
the semi-empirical ones than previous {\em ab initio} results employing the CPA
and/or spherical approximations, implying better consistency with thermal and
spin-wave data. Still, a significant difference exists, perhaps reflecting the
role of Hubbard correlations, which might reduce the effective values of $J$
via larger band gaps (recall that in the density functional calculations
the ferromagnetic case is metallic). In any case, we find both NN and NNN
interactions to be antiferromagnetic, consistent with the Goodenough-Kanamori
rules\cite{ref12,ref37} and experiment. We note that since these calculations
involve energy differences on the order of a few mRy, large {\bf k} point sets
were required in order to attain sufficient convergence. The above results were
calculated using 4096 special {\bf k} points in the full zone. Reduction of
this set to 1728 points yielded differences of less than 0.003 mRy and 0.001
mRy for $J_1$ and $J_2$, respectively, in LDA tests. Convergence with respect
to planewave cutoff was also checked. Increasing the cutoff from 18.9 Ry to
24.7 Ry yielded a difference of less than 0.002 mRy for both $J_1$ and $J_2$ in
LDA tests. Calculated Mn moments varied by less than 2.2\% over all magnetic
orderings in both LDA and GGA tests.

We deduce the distance dependence of the interactions from calculations on
distorted structures. For a small compression along [111], adjacent-plane NNs
move closer together while in-plane NNs move farther apart, so that $J_1
\rightarrow J_1'$ for adjacent-plane NNs while $J_1 \rightarrow J_1''$ for
in-plane NNs; while for a volume conserving strain NNN distances remain
unchanged to first order.\cite{ref25} Considering again a single formula unit,
the above Hamiltonian (\ref{eq1}) reduces to
\begin{equation}
H^{\rm FM} = (3J_1'+3J_1''+3J_2)S^2,
\end{equation}
\begin{equation}
H^{\rm AFI} = (-J_1'-J_1''+3J_2)S^2,
\end{equation}
and
\begin{equation}
H^{\rm AFII} = (-3J_1'+3J_1''-3J_2)S^2
\end{equation}
for FM, AFI, and AFII orderings, respectively. We deduce the exchange 
constants as before from the above expressions and {\em ab initio} total
energies corresponding to the appropriate magnetic orderings, at the
experimental volume and distortion. Our results are shown in Table~\ref{tab2},
along with semi-empirical values for comparison. As the table shows, our {\em
ab initio} results predict the same distance dependence as the semi-empirical
ones: whatever the nature of the NN interactions, whether direct or indirect,
their strength decreases with distance on the relevant scale, as may be
expected.

The observed compression along [111] now follows straightforwardly from the
above established distance dependence and antiferromagnetic nature of the
interactions. For a small compression angle $\delta > 0$,
\[ J_1' = J_1(1+c\delta), \]
\[ J_1'' = J_1(1-c\delta), \]
and, for a volume conserving strain, $J_2$ remains unchanged to first order;
where $J_1>0$ (antiferromagnetic NN interactions) and $c>0$ (strength decreases
with distance) as established above. Thus, for the AFII structure,
\[ H_{\rm strained}^{\rm AFII}-H_{\rm unstrained}^{\rm AFII} = 
   -6S^2 J_1 c\delta < 0; \]
and so the compression lowers energy. Physically, for a small compression, the
antiparallel-spin NNs move closer together while the parallel-spin NNs move
farther apart, while NNN distances remain unchanged. And so, since the NN
interaction is antiferromagnetic and decreases with distance, the energy is
lowered. Thus, while decisive in determining the magnetic ordering,\cite{ref14}
the strong NNN coupling through the oxygens has nothing (to first order) to do
with the observed rhombohedral distortion. Rather, it is the weaker
antiferromagnetic NN coupling which is decisive.

\section{Magnetic-order-induced charge anisotropy}
\label{magordindchargan}

The electric field gradient (EFG) tensor, the second derivative of the Coulomb
potential at the atomic position,\cite{ref38} provides a sensitive probe of the
electronic charge distribution; and due to this sensitivity, very accurate
distributions are required in order to calculate EFGs reliably. Blaha {\em et
al.}\cite{ref39,blaha96} were the first to show that the LAPW method could be
used to calculate accurate EFGs with no adjustable parameters. In subsequent
work\cite{ref40} it was shown that, due to the extreme sensitivity of the EFGs
to inaccuracies in the density, the local orbital extension of the method
(LAPW+LO),\cite{ref29} as we employ here, is sometimes required in order to
obtain reliable results.

Massidda {\em et al.}\cite{ref22} have investigated magnetic-order-induced 
anisotropies of two nonmagnetic (spin-independent) linear-response properties
of MnO, finding significant anisotropies in the zone-center optic phonon
frequencies and Born effective charge tensor despite the fact that the atomic
positions had cubic symmetry, so that the only noncubic ingredient was in the
spin channel. Since strong Hubbard correlations are thought to separate charge
and spin degrees of freedom in general, measurements of such couplings in
comparison with density-functional calculations can be particularly
illuminating. Here, we consider magnetic-order-induced anisotropy in a static,
nonmagnetic quantity: the charge density itself. Since the EFG (the
largest-magnitude eigenvalue of the EFG tensor) depends sensitively on the
charge density and vanishes for cubic symmetry, it provides a sensitive measure
of such anisotropy. And using this measure, our calculations show that magnetic
ordering can induce significant charge anisotropy in MnO---on the order of that
associated with noncubic crystal structure in some cases. Table~\ref{tab3}
shows our LDA and GGA results for MnO in various magnetic and structural
conformations. Cubic AFI and AFII structures show significant (up to $\approx
1.3 \times 10^{21}$ V/m$^2$) EFGs, indicating significant anisotropy due solely
to the magnetic ordering, with an order of magnitude difference in the Mn EFGs
in the AFI and AFII orderings. Generally speaking, GGA calculations are thought
to provide a better description than corresponding LDA results for the ground
state properties of materials which contain light elements and are not
strongly correlated. For example, GGA equilibrium lattice parameters and
structural coordinates are frequently in better agreement with experiment than
LDA results for such materials. However, while one may conjecture that this
should also be the case for EFG's, this is not well established yet because the
error bars on the experimental numbers are often larger than the LDA--GGA
differences due to the uncertainty in the nuclear quadrupole moments (note that
in most cases relative LDA--GGA differences are not as large as in the present
case). In fact, comparisons of calculated density functional EFG's with
experimental quadrupole splittings were even used to obtain a more accurate
value for the quadrupole moment of the Fe-57 nucleus.\cite{dufek95} In strongly
correlated systems, Coulomb correlations differentiate the orbitals more than
LDA calculations. In open shell materials, this generally would lead to larger
EFG's. For example, a strong effect of this type was demonstrated for the plane
Cu site in YBa$_2$Cu$_3$O$_7$.\cite{singh92} On the other hand, in closed shell
systems (like MnO), the effect may be opposite due to the suppression of charge
degrees of freedom relative to spin, i.e.\ a lowering of the EFG. The last row
of Table~\ref{tab3} corresponds to the AFII structure with the observed
rhombohedral distortion, and so constitutes our prediction for the observed
EFGs. These values should be amenable to experimental measurement and, as
mentioned, comparison with density-functional predictions should prove quite
useful.

\section{Summary and conclusions}  
\label{sumcon}

We have calculated the structural, electronic, and magnetic properties of MnO
from first principles, using the LAPW method, with both LDA and GGA
approximations to exchange and correlation. In both the LDA and GGA we found
the ground state to be of rhombohedrally distorted B1 structure with
compression along the [111] direction, antiferromagnetic with AFII ordering,
and insulating, consistent with experiment and previous HF and
density-functional calculations employing spherical, pseudopotential, and/or
strictly LDA approximations. The GGA was shown to yield a much improved lattice
constant. However, highly converged LDA and GGA results were shown to
overestimate the observed rhombohedral distortion, consistent with one recent
PWPP-GGA calculation but inconsistent with others, including recent LAPW-LDA
calculations. We showed that the distortion can be understood in terms of a
Heisenberg model with distance dependent NN and NNN interactions determined
from first principles, obtaining exchange constants in better agreement with
semi-empirical values than previous {\em ab initio} results employing the CPA
and/or spherical approximations, but still somewhat larger in general. Whereas
the NNN coupling plays the decisive role in the magnetic ordering, the weaker
NN coupling was shown to play the decisive role in the structural distortion.
Finally, we showed that magnetic ordering can induce significant charge
anisotropy, on the order of that associated with noncubic structure in some
cases, and gave density-functional predictions for EFGs in the observed
rhombohedrally distorted structure.

\acknowledgements

Support from the Office of Naval Research and National Research Council is
gratefully acknowledged.

\begin{figure}
\begin{center}
\epsfig{file=fig1.eps,width=0.4\linewidth,clip=true}
\end{center}
\caption{(a) AFI ([001]) and (b) AFII ([111]) magnetic orderings.}
\label{fig1}
\end{figure}

\begin{figure}
\begin{center}
\epsfig{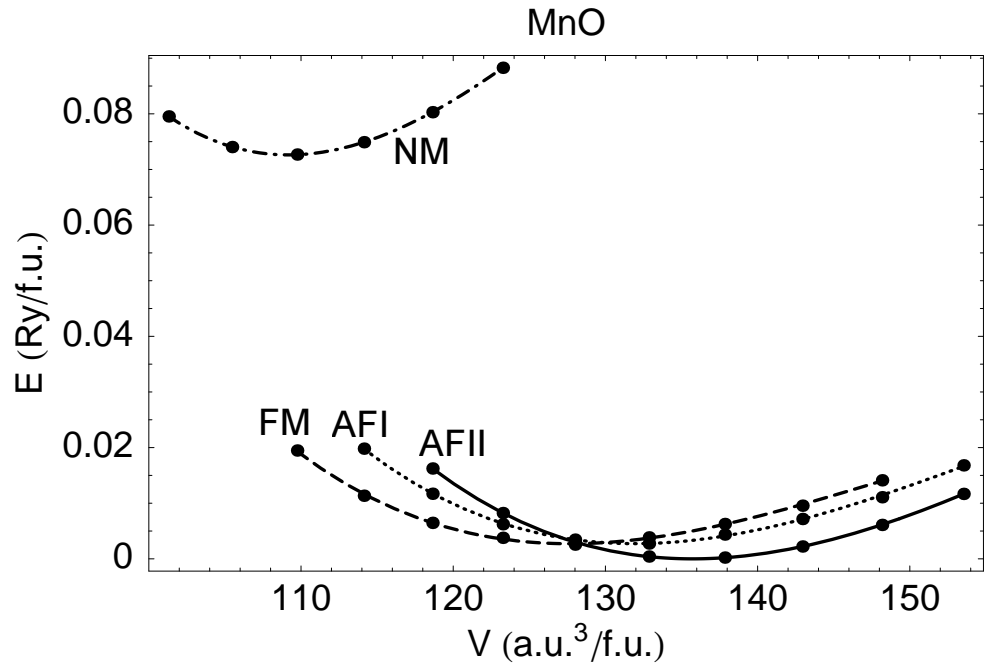}
\end{center}
\caption{LAPW-LDA results for energy vs.\ volume of MnO, predicting AFII
ground-state magnetic ordering.}
\label{fig2}
\end{figure}

\begin{figure}
\begin{center}
\epsfig{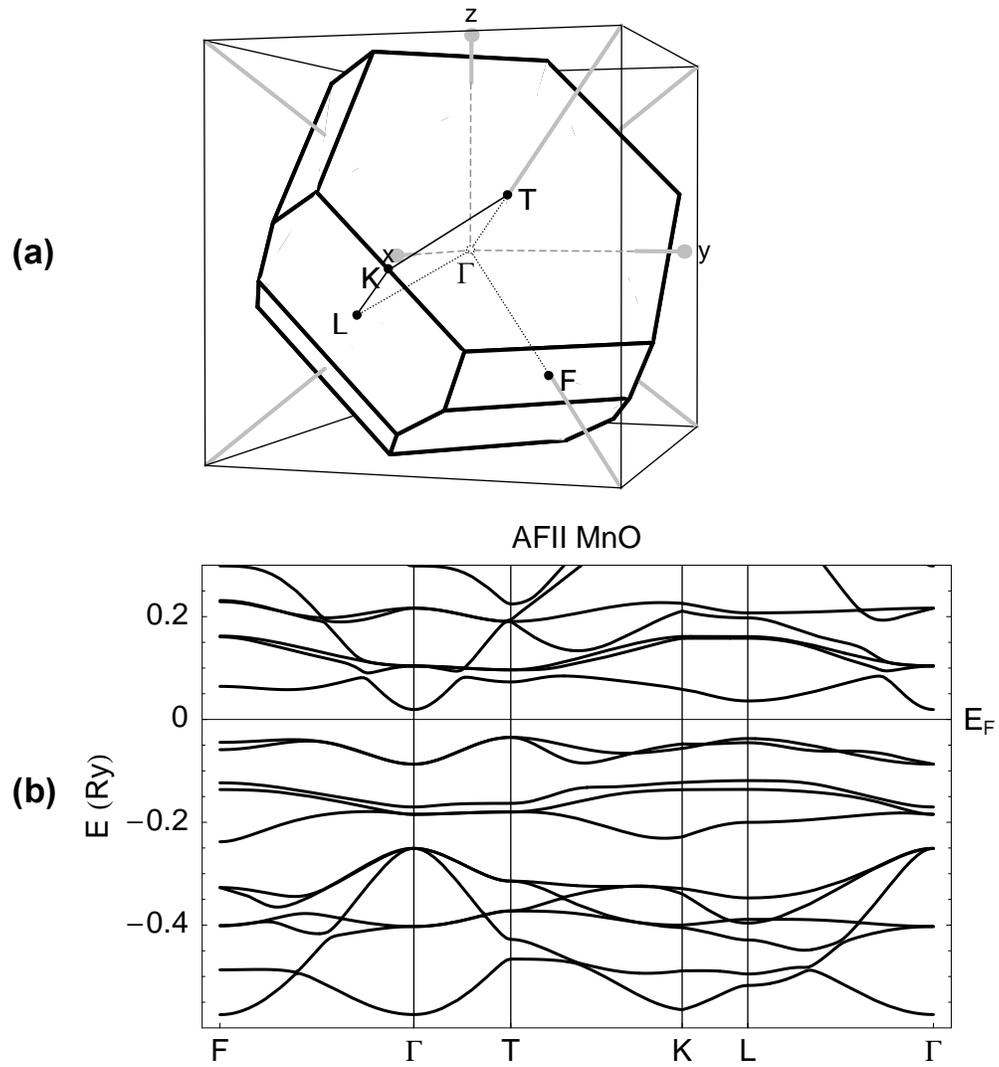}
\end{center}
\caption{(a) Rhombohedral Brillouin zone and (b) LAPW-LDA band structure of
AFII MnO.}
\label{fig3}
\end{figure}

\begin{figure}
\begin{center}
\epsfig{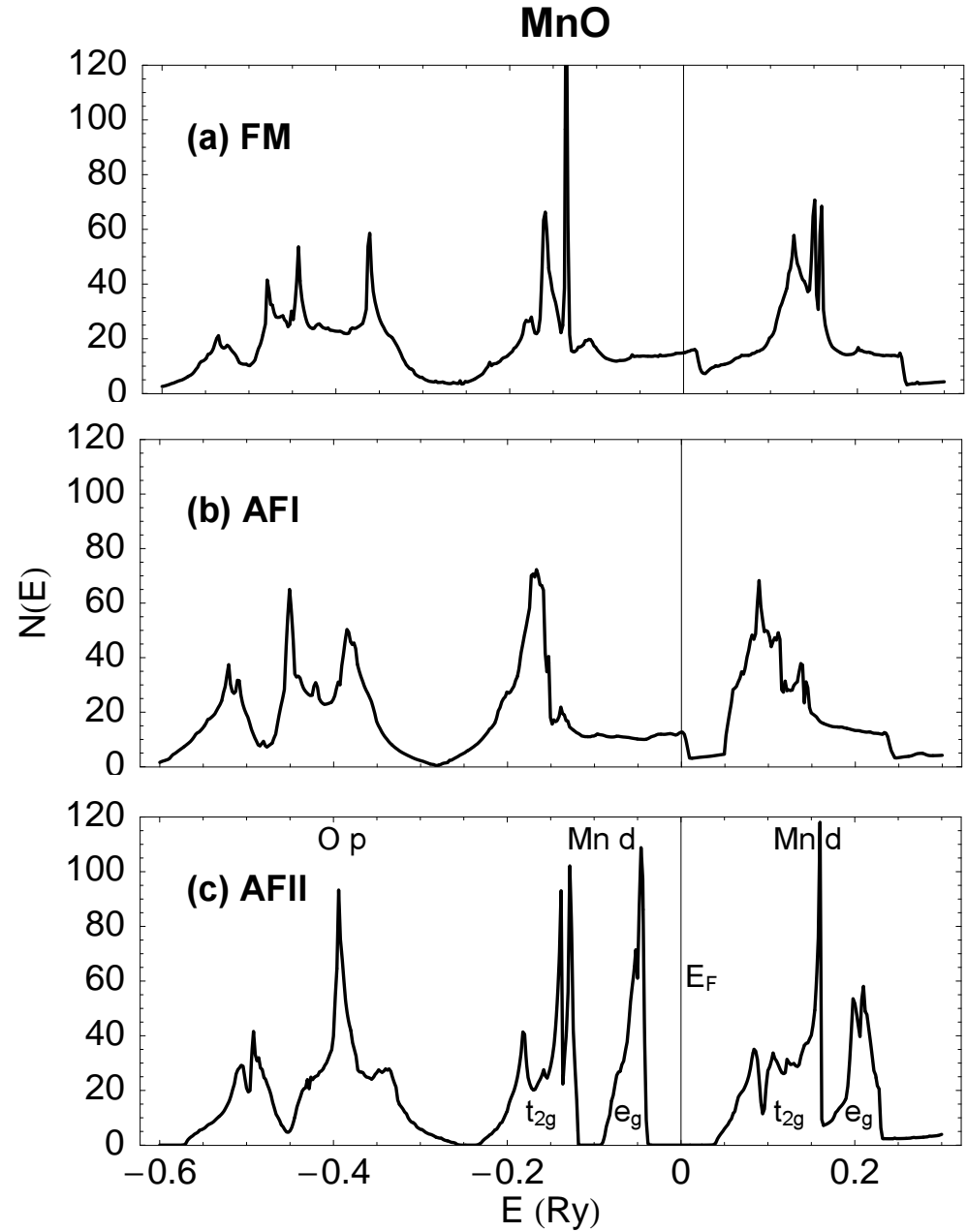}
\end{center}
\caption{LAPW-LDA density of states for (a) FM, (b) AFI, and (c) AFII MnO.
Only the AFII ordering is predicted to be insulating.}
\label{fig4}
\end{figure}

\begin{figure}
\begin{center}
\epsfig{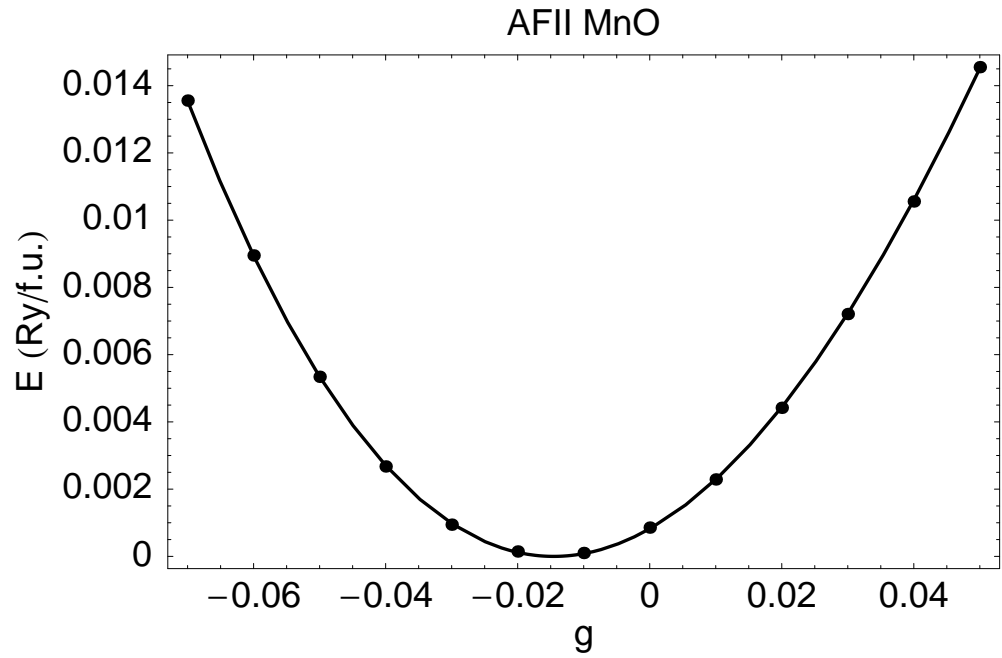}
\end{center}
\caption{LAPW-LDA energy vs.\ rhombohedral strain for AFII MnO at the
experimental volume. The minimizing strain of $g=0.01456$ corresponds to a
deviation of $1.68^\circ$ from the ideal cubic angle.}
\label{fig5}
\end{figure}

\begin{table}
\caption{LAPW LDA and GGA rhombohedral distortion angles for AFII MnO, and
comparison to previous calculations and experiment.}
\label{tab1}
\begin{tabular}{cccccc}
\multicolumn{2}{c}{LAPW} & PWPP-GGA & LCAO-HF & LAPW-LDA & EXPT \\
LDA & GGA 
   & (Ref.~\onlinecite{ref21}) & (Ref.~\onlinecite{ref18})
   & (Ref.~\onlinecite{ref22}) & (Ref.~\onlinecite{ref17}) \\
\hline
$1.68^\circ$ & $1.69^\circ$ 
   & $\approx 2.0^\circ$ & $0.47^\circ$ & $0.66^\circ$ & $0.62^\circ$
\end{tabular}
\end{table}

\begin{table}
\caption{Exchange constants (in K) for MnO based on LAPW LDA and GGA
calculations, and comparison to other {\em ab initio} and semi-empirical
results.}
\label{tab2}
\begin{tabular}{lddddddd}
   & \multicolumn{2}{c}{LAPW} & LMTO-ASA-LDA & ASW-LDA 
      & Semi-emp & Semi-emp & Semi-emp \\
   & LDA & GGA & (Refs.~\onlinecite{ref12,ref13}) & (Ref.~\onlinecite{ref27a})
      & (Ref.~\onlinecite{ref26}) & (Ref.~\onlinecite{ref35}) 
      & (Ref.~\onlinecite{ref36}) \\
\hline
$J_1$    &  9.8 & 18.8 & 24.5 & 30  & 10  &  8.9   & 8.5 \\
$J_2$    & 24.5 & 33.0 & 43.6 & 30  & 11  & 10.3   & 9.6 \\
$J_1'$   & 12.3 & 21.3 &      &     &     & 10.0   & 9.9 \\
$J_1''$  &  7.4 & 16.4 &      &     &     &  7.9   & 7.5
\end{tabular}
\end{table}

\begin{table}
\caption{LAPW LDA and GGA EFGs (in $10^{21}$ V/m$^2$) for various conformations
 of MnO. The last row corresponds to the observed rhombohedrally distorted
 ground state. The asymmetry parameters vanish in all cases.}
\label{tab3}
\begin{tabular}{lldd}
      &                 &  LDA   & GGA   \\
\hline
AFI:  & Mn              &  1.31  &  0.65 \\
      & O               &– 0.18  &  0.13 \\
\hline
AFII: & Mn              & -0.13  & -0.05 \\
      & O               &– 0.22  &  0.27 \\
\hline
AFII (distorted): & Mn  &– 0.10  &  0.20 \\
                  & O   & -0.12  & -0.08
\end{tabular}
\end{table}



\begin{references}
\bibitem{ref30}Sphere radii of 2.07 a.u. and 1.61 a.u. were used for Mn and O 
 atoms, repectively.
\end{references}
\end{document}